\documentstyle[12pt]{article}
\def\beq{\begin{equation}}
\def\eeq{\end{equation}}
\def\bea{\begin{eqnarray}}
\def\eea{\end{eqnarray}}
\def\bq{\begin{quote}}    
\def\eq{\end{quote}}

\def\bq{\begin{quote}}
\def\eq{\end{quote}}

\def\bq{\begin{quote}}
\def\eq{\end{quote}}

\def\gappeq{\mathrel{\rlap {\raise.5ex\hbox{$>$}}
{\lower.5ex\hbox{$\sim$}}}}

\def\lappeq{\mathrel{\rlap{\raise.5ex\hbox{$<$}}
{\lower.5ex\hbox{$\sim$}}}}

\def\bbz{fa Z \kern-8.9pt Z}

\begin{document}

\baselineskip 24pt
\newcommand{\sheptitle}
{Atmospheric and Solar Neutrinos from
Single Right-Handed Neutrino Dominance
and $U(1)$ Family Symmetry}

\newcommand{\shepauthor}
{S. F. King}

\newcommand{\shepaddress}
{Department of Physics and Astronomy,
University of Southampton, Southampton, SO17 1BJ, U.K.}

\newcommand{\shepabstract}
{We discuss a natural explanation
of both neutrino mass hierarchies 
{\it and} large neutrino mixing angles,
as required by the atmospheric neutrino data, in terms of
a single right-handed neutrino giving the
dominant contribution to the 23 block of the light effective
neutrino matrix, and illustrate this mechanism
in the framework of models with $U(1)$ family symmetries.
Sub-dominant contributions from other right-handed neutrinos are required to
give small mass splittings appropriate to the small angle MSW
solution to the solar neutrino problem.
We give general conditions for achieving this in the framework of
$U(1)$ family symmetry models
containing arbitrary numbers of right-handed neutrinos, and show how 
the resulting neutrino mass
hierarchies and mixing angles may be expanded in terms of the
Wolfenstein parameter.}

\begin{titlepage}
\begin{flushright}
hep-ph/9904210\\
\end{flushright}
\begin{center}
{\large{\bf \sheptitle}}
\bigskip \\ \shepauthor \\ \mbox{} \\ {\it \shepaddress} \\ \vspace{.5in}
{\bf Abstract} \bigskip \end{center} \setcounter{page}{0}
\shepabstract
\end{titlepage}

\section{Introduction}
There is now strong evidence for atmospheric neutrino oscillations
\cite{SK}, \cite{K} which confirms the earlier indications of the
effect \cite{early}.
The most recent analyses of Super-Kamiokande \cite{SK} involve
the hypothesis of
$\nu_{\mu}\rightarrow \nu_{\tau}$ oscillations
with maximal mixing $\sin^2 2\theta_{23} =1$
and a mass splitting of
$\Delta m_{23}^2 = 2.2\times 10^{-3}\ eV^2$.
Using all their data sets
analysed in different ways they quote $\sin^2 2\theta_{23} > 0.82$
and a mass splitting of
$5\times 10^{-4}\ eV^2 <\Delta m_{23}^2 < 6\times 10^{-3}\ eV^2$
at 90\% confidence level. 

The evidence for solar neutrino oscillations is almost as strong
\cite{SNU}.  There are
a panoply of experiments looking at different energy ranges, and
the best fit to all of them has been narrowed down to two basic
scenarios corresponding to either resonant oscillations
$\nu_e \rightarrow \nu_0$ (where for example 
$\nu_0$ may be a linear combination
of $\nu_{\mu} , \nu_{\tau}$) inside the Sun
(MSW \cite{MSW}) or ``just-so'' oscillations in the vacuum between the Sun
and the Earth \cite{justso1, justso2}.
There are three MSW fits and one vacuum oscillation fit:

(i) the small angle MSW 
solution is $\sin^2 2\theta_{12} \approx 5 \times 10^{-3}$
and $\Delta m_{12}^2 \approx  5\times 10^{-6}\ eV^2$;

(ii) the large angle MSW solution is $\sin^2 2\theta_{12} \approx 0.76$
and $\Delta m_{12}^2 \approx  1.8\times 10^{-5}\ eV^2$;

(iii) an additional MSW large angle solution exists with a lower probability 
\cite{BKS};

(iv) The vacuum oscillation solution is $\sin^2 2\theta_{12} \approx 0.75$
and $\Delta m_{12}^2 \approx  6.5\times 10^{-11}\ eV^2$ \cite{BKS}.

The standard model has zero neutrino masses, so any indication of
neutrino mass is very exciting since it represents
new physics beyond the standard model. In this paper
we shall assume the see-saw mechanism and no light sterile neutrinos.
The see-saw mechanism \cite{seesaw} implies that the 
three light neutrino masses
arise from some heavy ``right-handed neutrinos'' $N^p_R$ 
(in general there can be $Z$ gauge singlets with $p=1,\ldots Z$)
with a $Z\times Z$ Majorana mass matrix 
$M^{pq}_{RR}$ whose entries take values
at or below the unification scale $M_U \sim 10^{16}$ GeV.
The presence of electroweak scale Dirac mass terms $m_{LR}^{ip}$ 
(a $3 \times Z$ matrix) connecting the
left-handed neutrinos $\nu^i_L$ ($i=1,\ldots 3$)
to the right-handed neutrinos $N^p_R$ 
then results in a very light see-saw suppressed effective $3\times 3$ Majorana
mass matrix 
\beq
m_{LL}=m_{LR}M_{RR}^{-1}m_{LR}^T
\label{seesaw}
\eeq
for the left-handed neutrinos $\nu_L^i$, which are the light physical
degrees of freedom observed by experiment. 

Not surprisingly, following
the recent data,
there has been a torrent of theoretical papers concerned with
understanding how to extend the standard model in order to accomodate
the atmospheric and solar neutrino data \cite{torrent}. 
Perhaps the minimal extension of the standard model capable of accounting
for the atmospheric neutrino data involves
the addition of a {\em single} right-handed neutrino $N_R$ \cite{SK1, SK2}.
This is a special case 
of the general see-saw model with $Z=1$, so that
$M_{RR}$ is a trivial $1 \times 1$ matrix
and $m_{LR}$ is a $3 \times 1$ column matrix where
$m_{LR}^T=(\lambda_{\nu_e}, \lambda_{\nu_{\mu}},
\lambda_{\nu_{\tau}})v_2$ with
$v_2$ the vacuum expectation value of the Higgs field $H_2$
which is responsible for the neutrino Dirac masses,
and the notation for the Yukawa couplings $\lambda_i$
indicates that we are in
the charged lepton mass eigenstate basis $e_L, \mu_L, \tau_L$
with corresponding neutrinos $\nu_{e_L}, \nu_{\mu_L}, \nu_{\tau_L}$.
Since $M_{RR}$ is trivially invertible 
the light effective mass matrix in Eq.\ref{seesaw}
in the $\nu_{e_L}, \nu_{\mu_L}, \nu_{\tau_L}$ basis is simply given by
\beq
m_{LL}=\frac{m_{LR}m_{LR}^T}{M_{ RR}}
 =
\left( \begin{array}{lll}
 \lambda_{\nu_e}^2 &  \lambda_{\nu_e} \lambda_{\nu_{\mu}}
 & \lambda_{\nu_e} \lambda_{\nu_{\tau}}     \\
\lambda_{\nu_e} \lambda_{\nu_{\mu}} & \lambda_{\nu_{\mu}}^2 
& \lambda_{\nu_{\mu}} \lambda_{\nu_{\tau}} \\
\lambda_{\nu_e }\lambda_{\nu_{\tau}} 
& \lambda_{\nu_{\mu}} \lambda_{\nu_{\tau}} & \lambda_{\nu_{\tau}}^2
\end{array}
\right)\frac{v_2^2}{M_{RR}}.
\label{matrix}
\eeq
The matrix in Eq.\ref{matrix} has vanishing determinant which implies 
a zero eigenvalue. Furthermore the submatrix in the 23
sector has zero determinant which implies a second zero eigenvalue
associated with this sector.
In order to account for the Super-Kamiokande data we assumed \cite{SK1}:
\beq
\lambda_{\nu_e} \ll \lambda_{\nu_{\mu}} \approx \lambda_{\nu_{\tau}}.
\label{hierarchy}
\eeq
In the $\lambda_{\nu_e}=0$ limit the matrix in Eq.\ref{matrix} has zeros
along the first row and column, and so clearly
$\nu_e$ is massless, and the other two
eigenvectors are simply
\beq
\left(
\begin{array}{l}
\nu_0 \\
\nu_3
\end{array}
\right)
=
\left(
\begin{array}{ll}
c_{23} & -s_{23}\\
s_{23} & c_{23}
\end{array}
\right)
\left(
\begin{array}{l}
\nu_{\mu} \\
\nu_{\tau}
\end{array}
\right)
\label{bbasis}
\eeq
where $t_{23}=\lambda_{\nu_{\mu}}/\lambda_{\nu_{\tau}}$, with
$\nu_0$ being massless, due to the vanishing
of the determinant of the 23 submatrix and $\nu_3$ having a
mass $m_{\nu_3}=(\lambda_{\nu_{\mu}}^2 + \lambda_{\nu_{\tau}}^2)
v_2^2/M_{RR}$.
The Super-Kamiokande data is accounted for by choosing the parameters
such that
$t_{23} \sim 1$ and $m_{\nu_3} \sim 5\times 10^{-2}$ eV.
In this approximation 
the atmospheric neutrino data is then consistent with
$\nu_{\mu}\rightarrow \nu_{\tau}$ oscillations via two state mixing,
between $\nu_3$ and $\nu_0$. Note how the single right-handed neutrino
coupling to the 23 sector implies vanishing determinant of the 23
submatrix. This provides a natural explanation of
both large 23 mixing angles and a hierarchy
of neutrino masses in the 23 sector at the same time \cite{SK1}.

In order to account for the solar neutrino data a small mass perturbation
is required to lift the massless degeneracy
of the two neutrinos $\nu_0 ,\nu_e$. 
In our original approach \cite{SK1}
\footnote{
Another approach \cite{SK2}
which does not rely on additional right-handed neutrinos
is to use SUSY radiative corrections so that the one-loop corrected
neutrino masses are not zero but of order $10^{-5}$ eV suitable
for the vacuum oscillation solution.} we
introduced additional right-handed neutrinos in order to provide
a subdominant contribution to the effective mass matrix
in Eq.\ref{matrix}. To be precise we assumed a single dominant
right-handed neutrino below the unification scale, with additional
right-handed neutrinos at the unification scale which lead to 
subdominant contributions to the effective neutrino mass matrix.
By appealing to quark and lepton mass hierarchy we assumed
that the additional subdominant right-handed neutrinos generate 
a contribution 
$m_{\nu_{\tau}}\approx m_t^2/M_U \approx 2\times 10^{-3}$ eV,
where $m_t$ is the top quark mass.
The effect of this is to give a mass perturbation
to the 33 component of the mass matrix in Eq.\ref{matrix},
which results in $\nu_0 $ picking up a small mass, 
through its $\nu_{\tau}$ component, while $\nu_e $ remains
massless. Solar neutrino oscillations then arise from
$\nu_e \rightarrow \nu_0$ with the mass
splitting in the right range
for the small angle MSW solution, controlled by a small mixing angle
$\theta_{12} \approx 
\lambda_{\nu_e}/\sqrt{\lambda_{\nu_{\mu}}^2 + \lambda_{\nu_{\tau}}^2}$.
The main prediction of this scheme is of the neutrino oscillation
$\nu_e \rightarrow \nu_3$ with a mass difference 
$\Delta m_{13}^2 \approx \Delta m_{23}^2$
determined by the
Super-Kamiokande data and a
mixing angle $\theta_{13} \approx \theta_{12}$
determined by the small angle
MSW solution. Such oscillations may be observable at the proposed
long baseline experiments via $\nu_3 \rightarrow \nu_e$
which implies 
$\nu_{\mu} \rightarrow \nu_e$ oscillations
with $\sin^2 2\theta \approx 5 \times 10^{-3}$ (the small MSW angle)
and $\Delta m^2 \approx  2.2\times 10^{-3}\ eV^2$ (the
Super-Kamiokande square mass difference).

It should be clear from the foregoing discussion that the motivation for
single right-handed neutrino dominance (SRHND) is that
the determinant of the 23 submatrix of Eq.\ref{matrix}
approximately vanishes, leading to a natural explanation of {\em both} large
neutrino mixing angles {\em and} hierarchical neutrino masses in the 23
sector {\em at the same time} \cite{SK1}. Although the explicit example of
SRHND above was based on one of the right-handed neutrinos being lighter
than the others, it is clear that the idea of SRHND is more general
than this. In the present paper we shall define SRHND more generally
as the requirement that a single right-handed neutrino gives the
dominant contribution to the 23 submatrix of the light effective
neutrino mass matrix.
We shall propose SRHND as a general requirement and address
the following two questions:
\newline
1. What are the general conditions under
which SRHND in the 23 block can arise and 
how can we quantify the contribution of the sub-dominant 
right-handed neutrinos which are responsible
for breaking the massless degeneracy, and allowing the small angle MSW
solution? \newline
2. How can we understand the pattern of neutrino Yukawa couplings in
Eq.\ref{hierarchy} where the assumed equality 
$\lambda_{\nu_{\mu}} \approx \lambda_{\nu_{\tau}}$
is apparently at odds with the hierarchical Yukawa
couplings in the quark and charged lepton sector? \newline
In order to address the two questions above we shall discuss
SRHND in the context of a $U(1)$ family symmetry.
In fact neutrino masses and mixing
angles have already been studied in the context of $U(1)$
family symmetry models but in the models that exist to date
either SRHND is not present at all
\cite{Ross}, \cite{Ramond}, 
or where it is present its presence has apparently gone unnoticed
\cite{Altarelli}. 
\footnote{We should point out that the condition of the
approximately vanishing subdeterminant was first clearly stated
in ref.\cite{Altarelli}. However 
all the actual examples presented there correspond to 
a single right-handed neutrino giving the dominant contribution
to the 23 block of the effective neutrino mass matrix,
which is essentially the mechanism first proposed in ref.\cite{SK1}.
Also note that SRHND has very recently been applied to an
$SU(2)$ family symmetry model\cite{Barbieri}.}
Where there is no SRHND, either the contribution to the 23
mixing angles coming from the neutrino matrix are small \cite{Ross},
or the 23 neutrino mass hierarchy is not described by the Wolfenstein
expansion parameter \cite{Ramond}. Where the 23 neutrino
mass hierarchies are described by the Wolfenstein
expansion parameter {\em and} large 23 mixing angles naturally arise 
\cite{Altarelli}, we shall show that the physical reason why these
models are successful
is that a single right-handed neutrino is giving the dominant
contribution to the 23 submatrix of $m_{LL}$. 
We shall give general conditions that theories with $U(1)$ family
symmetry must satisfy in order to have SRHND and show that 
the models in \cite{Altarelli} satisfy these conditions.

\section{MSSM with $Z$ Right-handed Neutrinos}
To fix the notation, we assume the Yukawa terms of the minimal supersymmetric
standard model (MSSM) augmented by $Z$ right-handed neutrinos,
\bea
{\cal L}_{yuk}=\epsilon_{ab}\left[ -Y^u_{ij}H_u^aQ_i^bU^c_j
+Y^d_{ij}H_d^aQ_i^bD^c_j +Y^e_{ij}H_d^aL_i^bE^c_j
-Y^{\nu}_{ip}H_u^aL_i^bN^c_p + \frac{1}{2}M_{RR}^{pq}N^c_pN^c_q
\right] \nonumber \\
+H.c.
\label{MSSM}
\eea
where $\epsilon_{ab}=-\epsilon_{ba}$, $\epsilon_{12}=1$,
and the remaining notation is standard except that
the $Z$ right-handed neutrinos $N_R^p$ have been replaced by their
CP conjugates $N^c_p$ with $p,q=1,\dots, Z$.
When the two Higgs doublets get their vacuum expectation values
(VEVS) $<H_u^2>=v_2$, $<H_d^1>=v_1$ with $\tan \beta \equiv v_2 /v_1$
we find the terms
\beq
{\cal L}_{yuk}=v_2Y^u_{ij}U_iU^c_j
+v_1Y^d_{ij}D_iD^c_j +v_1Y^e_{ij}E_iE^c_j
+v_2Y^{\nu}_{ip}N_iN^c_p + \frac{1}{2}M_{RR}^{pq}N^c_pN^c_q +H.c.
\eeq
Replacing CP conjugate fields we can write in a matrix notation
\beq
{\cal L}_{yuk}=\bar{U}_Lv_2Y^uU_R
+\bar{D}_Lv_1Y^dD_R +\bar{E}_Lv_1Y^eE_R
+\bar{N}_Lv_2Y^{\nu}N_R + \frac{1}{2}N^T_RM_{RR}N_R +H.c.
\eeq
where we have assumed that all the masses and Yukawa couplings are
real and written $Y^\star =Y$.
The diagonal mass matrices are given by the following unitary transformations
\bea
v_2Y^u_{diag}=V_{uL}v_2Y^uV_{uR}^{\dag}={\rm diag (m_u,m_c,m_t)},\nonumber \\ 
v_1Y^d_{diag}=V_{dL}v_1Y^dV_{dR}^{\dag}={\rm diag(m_d,m_s,m_b)},\nonumber \\ 
v_1Y^e_{diag}=V_{eL}v_1Y^eV_{eR}^{\dag}={\rm diag(m_e,m_{\mu},m_{\tau})},
\nonumber \\ 
M_{RR}^{diag}=\Omega_{RR}M_{RR}\Omega_{RR}^{\dag}
={\rm diag(M_{R1},\ldots ,M_{RZ})},
\eea
where the unitary transformations are also orthogonal.
From Eq.\ref{seesaw} the light effective left-handed Majorana
neutrino mass matrix is
\beq
m_{LL}=v_2^2Y_{\nu}M_{RR}^{-1}Y_{\nu}^T
\label{mLL}
\eeq
Having constructed the light Majorana mass matrix it must then
be diagonalised by unitary transformations,
\beq
m_{LL}^{diag}=V_{\nu L}m_{LL}V_{\nu L}^{\dag}
={\rm diag(m_{\nu_1},m_{\nu_2},m_{\nu_3})}.
\eeq
The CKM matrix is given by
\beq
V_{CKM}=V_{uL}V_{dL}^{\dag}
\eeq
and its leptonic analogue is the MNS matrix \cite{MNS}
\beq
V_{MNS}=V_{\nu L}V_{eL}^{\dag}.
\eeq

\section{Wolfenstein Expansions}
The Wolfenstein parametrisation of the CKM matrix yields
the approximate form \cite{Wolf}:
\beq
V_{CKM} \sim
\left( \begin{array}{lll}
1 &  \lambda & \lambda^3 \\
\lambda & 1 & \lambda^2 \\
\lambda^3 & \lambda^2 & 1
\end{array}
\right)
\label{Wolf}
\eeq
where $\lambda \approx V_{us} \approx 0.22$.
The horizontal quark and lepton mass ratios may similarly
be expanded in terms of the Wolfenstein parameter:
\footnote{We follow the expansions in ref.\cite{Ramond}
even though $\frac{m_e}{m_{\tau}} \sim \lambda^5$ is a better fit.}
\beq
\frac{m_u}{m_t} \sim \lambda^8, \ \ 
\frac{m_c}{m_t} \sim \lambda^4, \ \ 
\frac{m_d}{m_b} \sim \lambda^4, \ \ 
\frac{m_s}{m_b} \sim \lambda^2, \ \ 
\frac{m_e}{m_{\tau}} \sim \lambda^4, \ \ 
\frac{m_{\mu}}{m_{\tau}} \sim \lambda^2. \ \ 
\eeq
Assuming the MSSM the vertical quark and lepton mass ratios
at $M_U$ are
\beq
\frac{m_b}{m_t} \sim \lambda^3, \ \ 
\frac{m_b}{m_{\tau}} \sim 1. \ \ 
\eeq
Assuming that $V_{CKM}\sim V_{uL} \sim V_{dL}$, and the diagonal
elements of the Yukawa matrices are of the same order as the
eigenvalues:\footnote{Again this is similar to ref.\cite{Ramond}
except that we allow a more general $\tan \beta$ dependence}
\beq
Y^u \sim
\left( \begin{array}{lll}
\lambda^8 &  \lambda^5 & \lambda^3 \\
-  & \lambda^4 & \lambda^2 \\
- & -  & 1
\end{array}
\right), \ \ \ \ 
Y^d \sim
\left( \begin{array}{lll}
\lambda^4 &  \lambda^3 & \lambda^3 \\
-  & \lambda^2 & \lambda^2 \\
- & -  & 1
\end{array}
\right)\lambda^n
\label{Ramond}
\eeq
where $\tan \beta \sim \lambda^{n-3}$.
Note that the CKM matrix only gives information about the upper
triangular parts of the quark Yukawa matrices.

The MNS matrix is less well determined, but Super-Kamiokande
tells us that $\theta_{23} \sim 1$ 
and the small angle MSW solution 
implies $\theta_{12} \sim \lambda^2$.
In addition for $\Delta m^2>9 \times 10^{-4} \ eV^2$ 
(i.e. over most of the atmospheric range) CHOOZ \cite{CHOOZ} 
fails to observe $\nu_e \rightarrow \nu_3$ and excludes
$\sin^2 2\theta_{13}>0.18$ or $\theta_{13}>0.22 $.
Hence CHOOZ allows $\theta_{13}\leq \lambda^2 $.
If we assume for the sake of argument that $\theta_{13}\sim \lambda^2 $
(recall that this is a prediction of SRHND which follows from
Eqs.\ref{matrix} and \ref{hierarchy})
then $V_{MNS}$ is given by:
\beq
V_{MNS} \sim
\left( \begin{array}{lll}
1 &  \lambda^2 & \lambda^2 \\
\lambda^2 & 1 & 1 \\
\lambda^2 & 1 & 1
\end{array}
\right)
\label{Wolflep}
\eeq
Then, in a similar way to the quarks,
assuming that $V_{MNS}\sim V_{\nu L} \sim V_{eL}$, and the diagonal
elements of the charged lepton matrix are of the same order as the
eigenvalues we deduce 
\beq
Y^e \sim
\left( \begin{array}{lll}
\lambda^4 &  \lambda^4 &  \lambda^2  \\
-  & \lambda^2 & 1 \\
- & -  & 1
\end{array}
\right) \lambda^n.
\label{Ramond'}
\eeq
The same argument applied to $m_{LL}$ runs into trouble because
the hierarchy between the second and third eigenvalues is 
apparently not consistent with $\theta_{23} \sim 1$.
To be precise Super-Kamiokande tells us that $m_{\nu_3} \approx 5 \times
10^{-2}$ eV, and small angle MSW tells us that
$m_{\nu_2} \approx 2 \times 10^{-3}$ eV, hence
\beq
\frac{m_{\nu_2}}{m_{\nu_3}}\sim \lambda^2.
\label{23hierarchy}
\eeq
The problem is how to generate such a hierarchy in the presence
of large neutrino mixing angles. Note that this problem
can be avoided for the charged lepton matrix in Eq.\ref{Ramond'}
due to the undetermined
32 element which can be small, but for the symmetric neutrino matrix
it is a problem. Fortunately the solution is provided by SRHND which
implies that $m_{LL}$
is given from Eqs.\ref{matrix} and \ref{hierarchy} as:
\beq
m_{LL} \sim
\left( \begin{array}{lll}
\lambda^4 &  \lambda^{2} &  \lambda^2  \\
\lambda^2  & 1 & 1 \\
\lambda^2 & 1  & 1
\end{array}
\right) m_{\nu_3}
\label{mLL''}
\eeq
It is clear that SRHND leads to the prediction
\beq
\frac{m_{\nu_1}}{m_{\nu_3}}\sim \lambda^4,
\eeq
in addition to the previously mentioned prediction $\theta_{13}\sim \lambda^2$.
The key to obtaining the hierarchy in Eq.\ref{23hierarchy}
from Eq.\ref{mLL''} is the requirement that the determinant of the
23 submatrix must vanish to order $\lambda^2$. Since this
subdeterminant naturally
vanishes for a single right-handed neutrino
coupling to the 23 sector, as in Eq.\ref{matrix}, all that is required
is for the subdominant right-handed neutrino to generate a
perturbation to the masses in the 23 sector which are of order
$\lambda^2$ smaller than the leading contribution. 
We shall now discuss how this can come about in the framework of
theories with broken $U(1)$ family symmetry.

\section{U(1) Family Symmetry}
The idea of accounting for the fermion mass spectrum via a broken
family symmetry has a long history \cite{FN}, \cite{textures}.
For definiteness we shall focus on a particular class of model based
on a single pseudo-anomalous $U(1)_X$ gauged family symmetry \cite{IR}.
We assume that the $U(1)_X$ is broken by the equal VEVs of two
MSSM singlets $\theta , \bar{\theta}$ which have vector-like 
charges $\pm 1$ \cite{IR}. 
Theories in which the $U(1)_X$ is broken by a chiral
MSSM singlet $\chi $ which has charge of one sign only, say $+1$,
have also been proposed \cite{shrock}, \cite{chiral}. In all these cases the 
$U(1)_X$ has anomalies in the effective low energy theory
below $M_U$ but these are compensated by string theory effects
at $M_U$ and the Green-Schwartz mechanism \cite{GS} provides a
dimension-five interaction term, whose structure demands a specific
pattern among the anomaly coefficients \cite{IR}:
\beq
A(SU(3)_c^2U(1)_X):A(SU(2)_L^2U(1)_X):A(U(1)_Y^2U(1)_X)=1:1:5/3
\label{anomalies}
\eeq

The $U(1)_X$ breaking scale is set by $<\theta >=<\bar{\theta} >$
where the VEVs arise from a Green-Schwartz computable
Fayet-Illiopoulos $D$-term which 
determines these VEVs to be one or two orders of magnitude
below $M_U$. Additional exotic 
matter which exists in vector-like pairs with opposite charges
$\pm X_i$  at a heavy
mass scale $M_V$ (generated by the VEVs of yet more
singlets) allows the Wolfenstein parameter to be generated by the
ratio \cite{IR}
\beq
\frac{<\theta >}{M_V}=\frac{<\bar{\theta} >}{M_V}= \lambda \approx 0.22
\label{expansion}
\eeq
The idea is that at tree-level the $U(1)_X$ family symmetry
only permits third family Yukawa couplings (e.g. the top quark
Yukawa coupling). Smaller Yukawa couplings are generated effectively
from higher dimension non-renormalisable operators corresponding
to insertions of $\theta$ and $\bar{\theta}$ fields and hence
to powers of the expansion parameter in Eq.\ref{expansion},
which we have identified with the Wolfenstein parameter.
The number of powers of the expansion parameter is controlled
by the $U(1)_X$ charge of the particular MSSM operator. 
\footnote{Of course this simple
picture may in reality be more complicated if several different
vector mass scales are assumed, and taking into account the order one
dimensionless couplings involving different $\theta$ and $\bar{\theta}$ 
fields coupling the MSSM fields to the heavy vector matter. 
By making various dynamical assumptions 
it is possible to generate several different expansion
parameters which may be in expanded non-integer powers \cite{Ross}.
It is also possible to introduce several $U(1)$ symmetries,
such as a model recently proposed based on a family-independent
pseudo-anomalous $U(1)_X$ symmetry together with two further anomaly-free
but family-dependent $U(1)$ symmetries \cite{Ramond}.
For our purposes here it is sufficient to assume 
a single $U(1)_X$ family symmetry with the single 
Wolfenstein expansion parameter in Eq.\ref{expansion} raised to
integer powers.}

The MSSM fields $Q_i$, $U^c_j$, $D^c_j$, $L_i$, $E^c_j$, $H_u$, $H_d$
are assigned $U(1)_X$ charges $q_i$, $u_j$, $d_j$, $l_i$, 
$e_j$, $h_u$, $h_d$ consistent with
Eq.\ref{anomalies}. This restricts the physical
values of the charges which we are permitted to assign.
\footnote{This restriction may be relaxed by assuming that the heavy vector
matter has $X_i$ charges chosen to cancel the anomalies, but we prefer
instead to regard this as a welcome constraint on the charges.
We shall, however, allow heavy MSSM singlets with arbitrary charges
to cancel $U(1)_X^3$ anomalies.}  
We do not impose any restriction 
on the $Z$ right-handed MSSM singlet neutrinos $N^c_p$ which therefore have 
unconstrained charges $n_p$. We shall
suppose that the right-handed neutrino Majorana mass matrix $M_{RR}$
arises from the VEV of another MSSM singlet $\Sigma$ with charge
$\sigma$ \cite{Ross}. 
The anomaly restriction means that there must exist a
physical basis where the Higgs charges are equal and opposite,
$h_u=-h_d$ in order to cancel their contributions to the anomalies,
and gives a zero charge to the $\mu H_u H_d$ term. 
The other operators in Eq.\ref{MSSM}
will in general have non-zero charges and 
from Eqs.\ref{expansion}, 
the associated Yukawa couplings and Majorana mass
terms may then be expanded in powers of the Wolfenstein parameter,
\bea
&& Y^u_{ij}\sim \lambda^{|q_i+u_j+h_u|}, \ \ 
Y^d_{ij}\sim \lambda^{|q_i+d_j+h_d|}, \ \ 
Y^e_{ij}\sim \lambda^{|l_i+e_j+h_d|}, \label{Yukexpch}\\
&& Y^{\nu}_{ip} \sim \lambda^{|l_i+n_p+h_u|}, \ \ 
M_{RR}^{pq} \sim \lambda^{|n_p + n_q + \sigma|}<\Sigma >. \ \ 
\label{Yukexpneut}
\eea

In the physical basis of charges discussed so far
the quarks and leptons must contribute to the anomalies
in the ratios in Eq.\ref{anomalies}. A corollary of this is that
the physical charges are related to traceless charges (denoted by
primes) by two flavour-independent $SU(5)$ shifts 
$\Delta t \equiv \Delta q =\Delta u = \Delta e$
and $\Delta f \equiv \Delta l =\Delta d $ \cite{IR}:
\beq
q_i '=q_i+\Delta t, \ \ u_i '=u_i+ \Delta t,\ \  
e_i '=e_i + \Delta t,\ \ 
l_i '=l_i + \Delta f, \ \ d_i '= d_i + \Delta f,
\label{primes}
\eeq
It is possible to absorb the $SU(5)$ shifts into the Higgs charges
by defining
\beq
h_u' \equiv h_u-2 \Delta t , \ \ h_d' \equiv h_d- \Delta t - \Delta f ,
\label{higgs'}
\eeq
so that
\beq
q_i+u_j+h_u=q_i'+u_j'+h_u', \ \ 
q_i+d_j+h_d=q_i'+d_j'+h_d', \ \ 
l_i+e_j+h_d=l_i'+e_j'+h_d'.
\label{equivalent}
\eeq
The couplings in Eq.\ref{Yukexpch} may then be equivalently expanded in terms
of primed charges.
Tracelessness implies that the first family charges may be eliminated
\beq
q_1'= -q_2'-q_3', \ \ u_1'=-u_2'-u_3', \ \ 
d_1'= -d_2'-d_3', \ \ l_1'=-l_2'-l_3', \ \ 
e_1'= -e_2'-e_3'.
\label{trace}
\eeq
Since the 33 component of the Yukawa matrices are either
renormalisable or related to $\tan \beta$ dependent integers
we can eliminate the primed Higgs charges using
\beq
h_u'=-q_3'-u_3', \ \ 
h_d'=n_d-q_3'-d_3'=n_e-l_3'-e_3'. \ \ 
\label{higgs}
\eeq
Using Eqs.\ref{equivalent}, \ref{trace}, \ref{higgs}
the Yukawa matrices in \ref{Yukexpch} may then be expressed as
$$
Y^u \sim
\left( \begin{array}{lll}
\lambda^{|\gamma_u+\delta_u|} & \lambda^{|\gamma_u+\beta_u|} 
& \lambda^{|\gamma_u|}\\
\lambda^{|\alpha_u+\delta_u|}& \lambda^{|\alpha_u+\beta_u|}
& \lambda^{|\alpha_u|}\\
\lambda^{|\delta_u|}& \lambda^{|\beta_u|} & 1
\end{array}
\right), \ \ \ \ 
Y^d \sim
\left( \begin{array}{lll}
\lambda^{|\gamma_d+\delta_d+n_d|} & \lambda^{|\gamma_d+\beta_d+n_d|} 
& \lambda^{|\gamma_d+n_d|}\\
\lambda^{|\alpha_d+\delta_d+n_d|}& \lambda^{|\alpha_d+\beta_d+n_d|}
& \lambda^{|\alpha_d+n_d|}\\
\lambda^{|\delta_d+n_d|}& \lambda^{|\beta_d+n_d|} & 
\lambda^{|n_d|} 
\end{array}
\right),
$$
\beq
Y^e \sim
\left( \begin{array}{lll}
\lambda^{|\gamma_e+\delta_e+n_e|} & \lambda^{|\gamma_e+\beta_e+n_e|} 
& \lambda^{|\gamma_e+n_e|}\\
\lambda^{|\alpha_e+\delta_e+n_e|}& \lambda^{|\alpha_e+\beta_e+n_e|}
& \lambda^{|\alpha_e+n_e|}\\
\lambda^{|\delta_e+n_e|}& \lambda^{|\beta_e+n_e|} & 
\lambda^{|n_e|} 
\end{array}
\right), \ \ \ \ 
\label{chargedYuks}
\eeq
where
\bea
&&\alpha_u=\alpha_d=q_2'-q_3', \ \ \alpha_e=l_2'-l_3', \nonumber \\
&&\beta_u=u_2'-u_3',\ \  \beta_d=d_2'-d_3', \ \ \beta_e=e_2'-e_3', \nonumber\\
&&\gamma_u=\gamma_d=-q_2'-2q_3',\ \  \gamma_e=-l_2'-2l_3', \nonumber \\
&&\delta_u=-u_2'-2u_3',\ \  \delta_d=-d_2'-2d_3', \ \ \delta_e=-e_2'-2e_3'.
\label{abcd}
\eea

The above analysis applies quite generally to any theory based
on a single pseudo-anomalous $U(1)_X$ gauged family symmetry.
However the quark and lepton charges
may be constrained by imposing unification constraints
on the theory. For example: 
\begin{itemize}
\item $SU(5)$ unification implies 
\footnote{Note that $SU(5)$ automatically guarantees Green-Schwartz
anomaly cancellation for any choice of charges.}
$l_i=d_i$, $q_i=u_i=e_i$
but allows $Z$ arbitrary right-handed neutrino charges $n_p$.
\item $SU(2)_R$ gauge symmetry implies that $Z=3$ with $n_i=e_i$
and $d_i=u_i$.
\item Left-right symmetry is stronger than $SU(2)_R$ and implies
$n_i=e_i=l_i$ and $d_i=u_i=q_i$.
\item Pati-Salam $SU(4)\times SU(2)_L \times SU(2)_R$ implies
$l_i=q_i$ and $u_i=d_i=e_i=n_i$.
\item $SO(10)$ unification implies $l_i=d_i=q_i=u_i=e_i=n_i$
\item Trinification $SU(3)^3$ implies $u_i=d_i$, $l_i=e_i=n_i$
and unconstrained $q_i$
\item Flipped $SU(5)\times U(1)$ implies $q_i=d_i=n_i$, $u_i=l_i$
and unconstrained $e_i$
\end{itemize}
As discussed in ref.\cite{Ross} 
these examples are difficult to reconcile with the data without
either appealing to group theoretical Clebsch relations or carefully
chosen dynamical assumptions. 
\footnote{The $SU(3)^3$ model discussed there looks the most natural.}
We shall therefore not impose such gauge
unification constraints here but instead consider the general case
in Eqs.\ref{chargedYuks}.

By comparing Eqs.\ref{chargedYuks},
to Eqs.\ref{Ramond}, \ref{Ramond'}
suitable choices of the integers $\alpha_a, \beta_a, \gamma_a, \delta_a$,
(where $a=u,d,e$) can readily be deduced.
Note especially that $m_b/m_{\tau} \sim 1$ implies
\beq
n=|n_e|=|n_d|, \ \ \tan \beta \sim \lambda^{n-3}.
\eeq
It is straightforward to scan over all the possible positive and negative
integers $\alpha_a, \beta_a, \gamma_a, \delta_a, n_a$ to find
acceptable Yukawa matrices from Eqs.\ref{chargedYuks}.
For example a special case is when
$\alpha_a, \beta_a, \gamma_a, \delta_a, n_a$ are all
positive definite integers \cite{Ramond}. In this case
from Eqs.\ref{Ramond}, \ref{Ramond'}, \ref{chargedYuks} we
find $\alpha_u=\alpha_d=2$, $\alpha_e=0$,
$\beta_u=2$, $\beta_d=0$, $\beta_e=2$,
$\gamma_u=\gamma_d=3$, $\gamma_e=2$,
$\delta_u=5$, $\delta_d=1$, $\delta_e=2$,
$n_e=n_d=n$.
\footnote{
Note that $n_e=n_d=n$ imposes the non-trivial constraint that
$\alpha_e+\beta_e+\gamma_e+\delta_e=\alpha_d+\beta_d+\gamma_d+\delta_d$
which is satisfied here. If for example we had taken 
$\frac{m_e}{m_{\tau}} \sim \lambda^5$ it would not be satisfied.}
The Yukawa matrices are then fully specified in this example,
up to a $\tan \beta$ dependence:
\beq
Y^u \sim
\left( \begin{array}{lll}
\lambda^8 &  \lambda^5 & \lambda^3 \\
\lambda^7 &  \lambda^4 & \lambda^2 \\
\lambda^5 &  \lambda^2 & 1
\end{array}
\right), \ \ \ \ 
Y^d \sim
\left( \begin{array}{lll}
\lambda^4 &  \lambda^3 & \lambda^3 \\
\lambda^3 & \lambda^2 & \lambda^2 \\
\lambda & 1 & 1
\end{array}
\right)\lambda^n, \ \ \ \ 
Y^e \sim
\left( \begin{array}{lll}
\lambda^4 &  \lambda^4 &  \lambda^2  \\
\lambda^2 & \lambda^2 & 1 \\
\lambda^2 & \lambda^2 & 1
\end{array}
\right) \lambda^n.
\eeq

Given $\alpha_a, \beta_a, \gamma_a, \delta_a, n_a$ above and using
Eqs.\ref{trace}, \ref{higgs}, \ref{abcd}
we find the following traceless:
\bea
&& q'_i=\frac{1}{3}(4,1,-5), \ \ u'_i=\frac{1}{3}(8,-1,-7), \ \
d'_i=\frac{1}{3}(2,-1,-1) \nonumber \\
&& l'_i=\frac{1}{3}(4,-2,-2), \ \ e'_i=\frac{1}{3}(2,2,-4),
\ \ h'_u=4, \ \ h'_d=2+n
\label{tracelesscharges}
\eea
The physical (unprimed) charges are by definition those which lead
to the Higgs charges satisfying $h_u=-h_d$.
Eq.\ref{higgs'} shows that there remains an ambiguity in the choice of
Higgs charges and hence in
$\Delta t , \Delta f$ which are two unknowns constrained by only
one relation, namely $-3\Delta t = h'_d +h'_u +\Delta f$.
We can regard $\Delta f$ as being a completely free
parameter whose choice specifies all the physical (unprimed) charges
uniquely. For example we may set the Higgs charges to be zero
by taking 
\footnote{Note that in
general both both $\Delta t$ and $\Delta f$ are non-zero
and so the family symmetry $U(1)_X$ cannot be anomaly-free and 
is instead pseudo-anomalous \cite{IR}.}
$\Delta t=-2$, $\Delta f = -n$ which enables
the physical (unprimed) charges to be deduced from Eq.\ref{primes}.
Other choices of $\Delta f$ will lead to different choices of physical
charges.

\section{SRHND and U(1) Family Symmetry}
We now turn our attention to the neutrino sector, which
is the main focus of this paper.
Since the $Z$ right-handed neutrinos are not constrained by anomaly
cancellation it is most convenient to work with physical (unprimed)
charges as in Eq.\ref{Yukexpneut}.
$Y^{\nu}$ clearly depends on the combination of lepton and Higgs charges
$$
l_i+ h_u =l'_i+h'_u/3-2h'_d/3-5\Delta f /3
$$
which is not fixed by
the primed charges due to the remaining freedom in $\Delta f$.
In dealing with the neutrino sector it is convenient to
absorb the Higgs charge $h_u$ into the
definition of the lepton charges $l_i$ so that Eq.\ref{Yukexpneut} becomes
\beq
Y^{\nu}_{ip} \sim \lambda^{|l_i+ n_p|}, \ \ 
M_{RR}^{pq} \sim \lambda^{|n_p + n_q + \sigma|}<\Sigma > \ \ 
\label{Yukexpneut'}
\eeq
where the redefined $l_i$ are related to the traceless charges $l'_i$
by arbitrary family-independent shifts, and
using Eq.\ref{tracelesscharges} may be written as:
\beq
l_i=(2+l_3,l_3,l_3)
\label{li}
\eeq
where the numerical value of $l_3$ remains a free choice.

The light Majorana matrix may then be constructed
from Eq.\ref{mLL} which we repeat below
\beq
m_{LL}=v_2^2Y_{\nu}M_{RR}^{-1}Y_{\nu}^T
\label{mLLagain}
\eeq
If we were to assume positive definite values for 
$l_i+n_p$ and $n_p + n_q + \sigma$ then the modulus signs
could be dropped and the right-handed neutrino charges $n_p$
would cancel when $m_{LL}$ is constructed 
from Eqs.\ref{mLLagain} and \ref{Yukexpneut'}  \cite{dropout}.
The argument relies on the observation that if the modulus signs
are dropped from Eq.\ref{Yukexpneut'} one can always write
\bea
&&Y^{\nu}=diag(\lambda^{l_1},\lambda^{l_2},\lambda^{l_3})Y_D
diag(\lambda^{n_1},\ldots ,\lambda^{n_Z}), \nonumber \\
&&M_{RR}=diag(\lambda^{n_1},\ldots ,\lambda^{n_Z})M_{M}
diag(\lambda^{n_1},\ldots ,\lambda^{n_Z})
\label{factors}
\eea
where $Y_D$ and $M_M$ are democratic matrices.
Inserting Eq.\ref{factors} into Eq.\ref{mLL} the right-handed
neutrino charges are seen to cancel. Such a cancellation
would imply that every right-handed neutrino would contribute
equally to every entry in $m_{LL}$ regardless of the right-handed
neutrino charges. From the point of view of SRHND it is therefore
important that such a cancellation does not take place, and so
we shall require that at least some of the combinations 
$l_i+n_p$ and $n_p + n_q + \sigma$ take negative values.
In such a case the choice of right-handed neutrino charges
will play an important role in determining $m_{LL}$,
and each particular choice of $n_p$ must be analysed
separately. 

At first sight the general case of $Z$ right-handed neutrinos
with unconstrained charges $n_p$ leading to non-positive
definite exponents in Eq.\ref{Yukexpneut'} seems to make the
determination of $m_{LL}$ an intractable problem.
However we have already argued that the atmospheric neutrino data
suggests SRHND in the 23 sector and this will lead to $m_{LL}$
of the form given in Eq.\ref{mLL''}. We shall now
formulate the general conditions which will lead to SRHND in the
23 sector. 

\subsection{One Right-handed Neutrino}
Let us first consider the case $Z=1$ where there is just a single
right-handed neutrino, which for later convenience
we shall refer to as $N^c_3$ with charge $n_3$.
In this case Eq.\ref{Yukexpneut'} becomes
\beq
Y^{\nu}_{i3} \sim \lambda^{|l_i+n_3|}, \ \ 
M_{RR}^{33} \sim \lambda^{|2n_3 + \sigma|}<\Sigma >. \ \ 
\label{Yukexpneut''}
\eeq
Being a $1 \times 1$ matrix $M_{RR}^{33}$ is trivially inverted
and we obtain from Eqs.\ref{mLLagain},
\beq
m_{LL}^{ij}\sim \lambda^{|l_i+n_3|}\lambda^{|l_j+n_3|}    
\frac{v_2^2}{M_{RR}^{33}}
\label{mLLSRHN}
\eeq
which should be compared to Eq.\ref{matrix}, where we identify
\footnote{Even though the couplings in Eq.\ref{matrix} were
defined in the diagonal charged lepton basis, the
identification is still valid to a consistent order of the expansion
parameter.}
\beq
Y^{\nu}_{i3} \sim \lambda^{|l_i+n_3|} \sim 
(\lambda_{\nu_e}, \lambda_{\nu_{\mu}}, \lambda_{\nu_{\tau}})
\eeq
Then Eq.\ref{mLL''} requires that
\beq
|l_2+n_3|=|l_3+n_3|, \ \ \ \ |l_1+n_3|-|l_3+n_3|=2
\label{modulus}
\eeq
If both $l_2+n_3$ and $l_3+n_3$ have the same sign (SS) then
$l_2=l_3$, whereas if they have opposite signs (OS) then
$l_2+l_3=-2n_3$. Similarly 
if both $l_1+n_3$ and $l_3+n_3$ have the SS then
$l_1-l_3=2$, whereas if they have OS then
$l_1+l_3=2-2n_3$. 
Interestingly the SS cases $l_2=l_3$, $l_1-l_3=2$ 
have already arisen in the example in Eq.\ref{tracelesscharges}, 
which corresponds to $l_i$ charges in Eq.\ref{li}.
This is no surprise since it originates from the 
charged lepton Yukawa matrix in
Eq.\ref{Ramond'} which follows from the assumption
$V_{MNS}\sim V_{\nu L} \sim V_{eL}$ and the Super-Kamiokande
data and the MSW solution. 

To summarise,
from Eqs.\ref{Yukexpneut''}, \ref{mLLSRHN} and imposing Eq.\ref{modulus}
the single right-handed neutrino included so far leads to
\beq
m_{LL} \sim
\left( \begin{array}{lll}
\lambda^4 &  \lambda^{2} &  \lambda^2  \\
\lambda^2  & 1 & 1 \\
\lambda^2 & 1  & 1
\end{array}
\right) m_{\nu_3}
\label{mLL'''}
\eeq
where the atmospheric neutrino mass is given 
\beq
m_{\nu_3}\sim \lambda^{2|l_3+n_3|-|2n_3 +\sigma |}       
\frac{v_2^2}{<\Sigma >}
\eeq
With only a single right-handed neutrino $m_{LL}$ in Eq.\ref{mLL'''}
has two zero eigenvalues, and a vanishing determinant of the 23
submatrix, as in Eq.\ref{matrix}.
In order to implement the small angle
MSW solution we need to include the effect of subdominant right-handed
neutrinos which break the massless degeneracy. 
SRHND requires that the elements in the
23 sector of Eq.\ref{mLL'''} must receive corrections of order
$\lambda^2$ from the subdominant neutrinos so that the
determinant of the 23 submatrix only {\em approximately}
vanishes to this order
leading to a small eigenvalue of order $\lambda^2$ and
the desired mass hierarchy in Eq.\ref{23hierarchy}.

\subsection{Two Right-handed Neutrinos}
We now include a second right-handed neutrino $N_2^c$ with charge
$n_2$, in addition to $N_3^c$ with charge $n_3$.
With two right-handed neutrinos, $Z=2$,
the heavy Majorana mass matrix from
Eq.\ref{Yukexpneut'} is
\beq
M_{RR} \sim
\left( \begin{array}{ll}
\lambda^{|2n_2+\sigma|} & \lambda^{|n_2+n_3+\sigma|} \\
\lambda^{|n_2+n_3+\sigma|} & \lambda^{|2n_3+\sigma|}
\end{array}
\right) <\Sigma >
\label{mRR2}
\eeq
For SRHND we clearly require $n_2 \neq n_3$ to avoid the two
right-handed neutrinos contributing democratically.
More generally for SRHND we need to avoid large right-handed neutrino mixing
angles. 
If we assume without loss of generality that 
$\lambda^{|2n_2+\sigma|} > \lambda^{|2n_3+\sigma|}$, so that
$N_2^c$ is heavier than $N_3^c$, then this implies
\beq
|2n_2+\sigma| < |2n_3+\sigma|
\label{c1}
\eeq
Then the small mixing angle requirement is
\beq
|2n_2+\sigma| < |n_2+n_3+\sigma| 
\label{c2}
\eeq
The lightest eigenvalue is of order the diagonal element provided
\beq
|2n_2+\sigma| \leq  2|n_2+n_3+\sigma|- |2n_3+\sigma| 
\label{c3}
\eeq
Assuming all these conditions are met then $M_{RR}$ will be 
diagonalised by small angle rotations and have hierarchical
eigenvalues set by the diagonal elements. 
As a first approximation we may drop the off-diagonal
elements and write 
\beq
M_{RR}\approx diag(M_{R2}, M_{R3})
\label{diag}
\eeq
where
\beq
M_{R2} \sim \lambda^{|2n_2+\sigma|}<\Sigma >, 
\ \ M_{R3} \sim \lambda^{|2n_3+\sigma|}<\Sigma >
\label{MR23}
\eeq
Then the light Majorana matrix is given by adding the separate contribution
from each of the two right-handed neutrinos
\beq
m_{LL}^{ij}=v_2^2\left( \frac{{{Y}}_{\nu}^{i2} {{Y}}_{\nu}^{j2 } }{M_{R2}}+
\frac{{{Y}}_{\nu}^{i3} {{Y}}_{\nu}^{j3} }{M_{R3}} \right)
\eeq
It is clear that the dominant
contribution to a particular element of $m_{LL}$ will come from
the right-handed neutrino which is at the same time the
lightest, and couples the most strongly to left-handed neutrinos.
Without loss of generality we have taken $N_3^c$ to be the lighter
right-handed neutrino and to give the dominant contribution to the
23 block of $m_{LL}$ in Eq.\ref{mLLSRHN}. We therefore write the
subdominant contribution coming from the second right-handed
neutrino $N_2^c$ as
\beq
\delta m_{LL}^{ij}                            
=\lambda^{|l_i+n_2|}\lambda^{|l_j+n_2|} \frac{v_2^2}{M_{R2}}
\eeq
As discussed below Eq.\ref{mLL'''} we require:
\beq
\frac{\delta m_{LL}^{33}}{m_{LL}^{33}} = 
\frac{\lambda^{2|l_3+n_2|} }{\lambda^{2|l_3+n_3|} }
\frac{M_{R3}}{M_{R2}} \sim \lambda^2
\label{cond}
\eeq
From Eqs.\ref{MR23},
\beq
\frac{M_{R3}}{M_{R2}} \sim 
\lambda^{|2n_3+\sigma|-|2n_2+\sigma|}
\eeq
so Eq.\ref{cond} implies the condition
\beq
2|l_3+n_2|-2|l_3+n_3|+|2n_3+\sigma|-|2n_2+\sigma|=2
\label{condition}
\eeq
We already observed that the required MSW perturbation is
\beq
\delta m_{LL}^{33} \sim \frac{v_2^2}{M_U}
\eeq
so we deduce
\beq
\frac{M_{R2}}{M_U}\sim \lambda^{2|l_3+n_2|}, \ \ 
\frac{<\Sigma >}{M_U}\sim \lambda^{2|l_3+n_2|- |2n_2+\sigma|}
\eeq

There is the further requirement that the powers of $\lambda$
occuring in $M_{RR}$ and $Y_{\nu}$ be either integer or half-integer.
\footnote{In the case of half-integer powers this implies that
the $\theta$, $\bar{\theta}$ fields which break the $U(1)_X$ symmetry
must have charges $\pm 1/2$ and the expansion parameter in
Eq.\ref{expansion} must be redefined so that
$\frac{<\theta >}{M_V}=\frac{<\bar{\theta} >}{M_V}= \lambda^{1/2}$,
as in ref.\cite{Ross}.} By scanning over half-integer and integer
values of $l_3,n_2,n_3,\sigma$ we find that there are no solutions
which satisfy all the above constraints for integer powers of $\lambda$ 
in $M_{RR}$ and $Y_{\nu}$.\footnote{I am grateful to
Y. Nir (private communication) for pointing this out.} 
However there are a large number
of solutions involving half-integer powers of $\lambda$ 
in $M_{RR}$ and $Y_{\nu}$ (of course $m_{LL}$ in Eq.\ref{mLL''}
always involves integer powers of $\lambda$.)
The condition in Eq.\ref{condition} may be achieved in various ways
with $N_3^c$ being lighter than $N_2^c$ by a factor,
$M_{R3}/M_{R2} \sim \lambda^{|2n_3+\sigma|-|2n_2+\sigma|}\sim \lambda^a$,
and the ratio of the Dirac couplings of $N_2^c$, $N_3^c$ to $L_3$
given by $\lambda^{{2|l_3+n_2|}- 2|l_3+n_3|} \sim
\lambda^{2-a}$, where $a>0$ is a positive integer.
For example $l_3=-1/2,n_2=0,n_3=1,\sigma =0$ satisfies all the
conditions with $a=2$ and $Y_{\nu}$ involving half-integer exponents.
Further examples are listed in Table 1.

\begin{table}[tbp]
\hfil
\begin{tabular}{ccccccc}
\hline
           $l_3$  & $n_2$& $n_3$ & $\sigma$ &   $a$   
\\ \hline
             -1   & -1/2  & 0    & 1 &  1 \\ \hline
             -1   & 0  & 1/2    & 0 &  1 \\ \hline
             -1   & 0  & 1/2    & 1/2 &  1 \\ \hline
             -1   & 0  & 1/2    & 1 &  1 \\ \hline
             -1   & 1/2  & 1    & -1 &  1 \\ \hline
             -1   & 1/2  & 1    & -1/2 &  1 \\ \hline
             -1   & 1/2  & 1    & 0 &  1 \\ \hline
             -1   & 1/2  & 1    & 1/2 &  1 \\ \hline
             -1   & 1/2  & 1    & 1 &  1 \\ \hline
             -1/2   & -1/2  & 0    & 1 & 1  \\ \hline
              -1/2  & 0  & 1/2    & 0 &  1 \\ \hline
              -1/2  & 0  & 1/2    & 1/2 & 1  \\ \hline
              -1/2  &  0 &  1/2   & 1  & 1  \\ \hline            
               0 & -1/2  &  0   & 1 &  1 \\ \hline
               0 & 1/2  & 0    & -1 &  1 \\ \hline
               1/2 & 0  & -1/2    & -1 & 1  \\ \hline
               1/2 & 0  & -1/2    & -1/2 & 1  \\ \hline
               1/2 & 0  & -1/2    & 0 & 1  \\ \hline
               1/2 & 1/2  & 0    & -1 & 1  \\ \hline
               1 & -1/2  & -1    & -1 & 1  \\ \hline
               1 & -1/2  & -1    & -1/2 & 1  \\ \hline
               1 & -1/2  & -1    & 0 & 1  \\ \hline
               1 & -1/2  & -1    & 1/2 & 1  \\ \hline
               1 & -1/2  & -1    & 1 & 1  \\ \hline
               1 & 0  & -1/2    & -1 & 1  \\ \hline
               1 & 0  & -1/2    & -1/2 & 1  \\ \hline
               1 & 0  & -1/2    & 0 & 1  \\ \hline
               1 & 1/2  & 0    & -1 & 1  \\ \hline     
              -1/2  &  0 &  1   & 0 &  2 \\ \hline
               -1/2 &  0 &  1   & 1/2 & 2  \\ \hline
               -1/2 &  0 &  1   & 1 &  2 \\ \hline
               0 & -1/2  &  1/2   & 1 &  2 \\ \hline             
               0 & 1/2  & -1/2    & -1 & 2  \\ \hline
               1/2 & 0  & -1    & -1 & 2  \\ \hline
               1/2 & 0  & -1    & -1/2 & 2  \\ \hline
               1/2 & 0  & -1    & 0 & 2  \\ \hline
               0 &  -1/2 & 1    & 1 &  3 \\ \hline
               0 &  1/2 & -1    & -1 & 3  \\ \hline
\end{tabular}
\hfil
\caption{\footnotesize 
Simple $Z=2$ examples which satisfy all the conditions of SRHND 
given in the text.}
\end{table}

\subsection{Three Right-handed Neutrinos}
We now wish to extend the discussion to include three right-handed
neutrinos $Z=3$, by introducing a third right-handed neutrino
$N_1^c$ with charge $n_1$ in addition to the 
two already introduced above. Again we shall suppose that $N_3^c$
gives the dominant contribution to the 23 sector masses.
As for the $Z=2$ case we require $n_3 \neq n_2,n_1$,
and we need to ensure that $N_3^c$ does not have large mixing angles
in $M_{RR}$ in order to isolate it from the other right-handed
neutrinos. 
This can be ensured by a sequence of conditions
similar to Eqs.\ref{c1}, \ref{c2}, \ref{c3}. 
Then, after small angle rotations, $M_{RR}$ can be written in 
block diagonal form.
\beq
M_{RR} \sim
\left( \begin{array}{lll}
\lambda^{|2n_1+\sigma|} & \lambda^{|n_1+n_2+\sigma|} & 0 \\
\lambda^{|n_2+n_1+\sigma|} & \lambda^{|2n_2+\sigma|} & 0 \\
0  & 0  & \lambda^{|2n_3+\sigma|} \\
\end{array}
\right) <\Sigma >
\label{mRR3}
\eeq
which is the analogue of Eq.\ref{diag}.
The new feature of the $Z=3$ case compared to the $Z=2$ case is that
there are now several possibilities for the structure of the upper
$2 \times 2$ block in Eq.\ref{mRR3} which are all consistent with
SRHND, which are listed below.

``Diagonal dominated'' 
corresponding to $|n_1+n_2+\sigma|>min(|2n_1+\sigma|,|2n_2+\sigma|)$:
\beq
M_{RR}^{upper} \sim
\left( \begin{array}{ll}
\lambda^{|2n_1+\sigma|} & 0 \\
0 & \lambda^{|2n_2+\sigma|}
\end{array}
\right) <\Sigma >
\label{mRR2a}
\eeq

``Off-diagonal dominated''
corresponding to $|n_1+n_2+\sigma|<|2n_1+\sigma|,|2n_2+\sigma|$:
\beq
M_{RR}^{upper} \sim
\left( \begin{array}{ll}
0 & \lambda^{|n_1+n_2+\sigma|} \\
\lambda^{|n_2+n_1+\sigma|} & 0 
\end{array}
\right) <\Sigma >
\label{mRR2b}
\eeq

``Democratic'' corresponding to 
$|n_1+n_2+\sigma| = |2n_1+\sigma| = |2n_2+\sigma|$:
\beq
M_{RR}^{upper} \sim
\left( \begin{array}{ll}
\lambda^{|2n_1+\sigma|} & \lambda^{|n_1+n_2+\sigma|} \\
\lambda^{|n_2+n_1+\sigma|} & \lambda^{|2n_2+\sigma|} 
\end{array}
\right) <\Sigma >
\label{mRR2c}
\eeq

In the ``diagonal dominated'' case after small angle rotations
the light effective Majorana mass matrix in Eq.\ref{mLLagain}
may be calculated in the diagonal right-handed neutrino basis
\beq
m_{LL}=v_2^2Y_{\nu}M_{RR}^{-1}Y_{\nu}^T
=v_2^2Y_{\nu}\Omega_{RR}^{\dag}(M_{RR}^{diag})^{-1}
\Omega_{RR}Y_{\nu}^T
\label{mLL''''}
\eeq
The advantage of working in a diagonal right-handed neutrino 
mass basis is that $(M_{RR}^{diag})^{-1}
={\rm diag(M_{R1}^{-1},M_{R2}^{-1},M_{R3}^{-1})}$ so
if we define ${\tilde{Y}}_{\nu} \equiv Y_{\nu}\Omega_{RR}^{\dag}$ as the
neutrino Yukawa matrix in the diagonal right-handed neutrino basis,
then the effective light mass matrix elements are given from 
Eq.\ref{mLL''''} by 
\beq
m_{LL}^{ij}=\sum_{p=1}^3
v_2^2\frac{{\tilde{Y}}_{\nu}^{ip} {\tilde{Y}}_{\nu}^{jp } }{M_{Rp}}
\label{mLL'}
\eeq
In this case $\Omega_{RR}$ involves small angle rotations
and so $\tilde{Y}_{\nu} \sim Y_{\nu}$, and the contributions to $m_{LL}$
from the neutrinos $N_1^c,N_2^c$ are:
\beq
\delta m_{LL}^{ij}=
v_2^2\left( \frac{{{Y}}_{\nu}^{i1} {{Y}}_{\nu}^{j1 } }{M_{R1}}+
\frac{{{Y}}_{\nu}^{i2} {{Y}}_{\nu}^{j2} }{M_{R2}} \right)
\eeq
where
\beq
M_{R1} \sim \lambda^{|2n_1+\sigma|}<\Sigma >, 
\ \ M_{R2} \sim \lambda^{|2n_2+\sigma|}<\Sigma >
\label{MR1MR2}
\eeq
and from Eq.\ref{Yukexpneut'}
$Y_{\nu}^{ip}=\lambda^{|l_i+n_p|}$.
Similar to Eq.\ref{cond} in this case we require
\beq
\frac{\delta m_{LL}^{33}}{m_{LL}^{33}} \sim
\frac{\lambda^{2|l_3+n_1|} }{\lambda^{2|l_3+n_3|} }
\frac{M_{R3}}{M_{R1}}
+
\frac{\lambda^{2|l_3+n_2|} }{\lambda^{2|l_3+n_3|} }
\frac{M_{R3}}{M_{R2}}
\sim \lambda^2
\label{conda}
\eeq
Thus the conditions for the ``diagonal dominated case'' are:
\bea
&& 2|l_3+n_1|-2|l_3+n_3| + |2n_3+\sigma| -|2n_1+\sigma|\geq 2,
\nonumber \\
&& 2|l_3+n_2|-2|l_3+n_3|+ |2n_3+\sigma|-|2n_2+\sigma|  \geq 2
\label{a}
\eea
where at least one of the inequalities must be saturated.

In the ``off-diagonal dominated'' case $M_{RR}$ can again be
simply inverted leading to
\beq
\delta m_{LL}^{ij}=
v_2^2\left( \frac{{{Y}}_{\nu}^{i1} {{Y}}_{\nu}^{j2 } }{M_{R12}}+
\frac{{{Y}}_{\nu}^{i2} {{Y}}_{\nu}^{j1} }{M_{R12}}
\right)
\eeq
where
\beq
M_{R12} \sim \lambda^{|n_1+n_2+\sigma|}<\Sigma >. 
\label{MR12}
\eeq
Again similar to Eq.\ref{cond} we require
\beq
\frac{\delta m_{LL}^{33}}{m_{LL}^{33}} \sim
\frac{\lambda^{|l_3+n_1|+|l_3+n_2|} }{\lambda^{2|l_3+n_3|} }
\frac{M_{R3}}{M_{R12}}
\sim \lambda^2
\label{condb}
\eeq
Thus the condition for the ``off-diagonal dominated case'' is:
\beq
|l_3+n_1|+|l_3+n_2|-2|l_3+n_3| + |2n_3+\sigma| -|n_1+n_2+ \sigma|= 2.
\label{b}
\eeq

In the ``democratic'' case 
$M_{RR}$ can be readily inverted leading to a result of order
\beq
\delta m_{LL}^{ij}\sim
v_2^2\left( \frac{{{Y}}_{\nu}^{i1} {{Y}}_{\nu}^{j1 } }{M}\right)
\eeq
where the right-handed neutrino masses in the upper block, $M$, are all equal
by the democratic assumption and we have specialised to
$n_1=n_2$ which implies from Eq.\ref{Yukexpneut'} that
${Y}_{\nu}^{i1}\sim {Y}_{\nu}^{i2}$.
Once again similar to Eq.\ref{cond} we require
\beq
\frac{\delta m_{LL}^{33}}{m_{LL}^{33}} \sim
\frac{\lambda^{2|l_3+n_1|} }{\lambda^{2|l_3+n_3|} }
\frac{M_{R3}}{M}
\sim \lambda^2
\label{condc}
\eeq
Thus the condition for the ``democratic case'' is:
\beq
2|l_3+n_1| -2|l_3+n_3| + |2n_3+\sigma| -|2n_1+ \sigma|= 2.
\label{c}
\eeq

In practice examples of all three kinds 
can easily be constructed along the same lines as the
explicit $Z=2$ case. The ``democratic'' case with $n_1=n_2$ 
is isomorphic to the $Z=2$ case. 
The $Z=2$ results trivially generalise in this case to
$n_p=(n_2,n_2,n_3)$ where some examples of
charges were listed in Table 1. For example
$l_3=-1/2, n_p=(0,0,1), \sigma =0$ satisfies all the ``democratic''
conditions with $a=2$ and $Y_{\nu}$ involving half-integer exponents.
Clearly in the ``democratic'' case the $Z=2$ results can immediately
be generalised to any number of right-handed neutrinos $Z$ with
$n_p=(n_2, \ldots, n_2,n_3)$, where $(n_2, n_3)$ are the $Z=2$ charges.

The ``diagonal dominated'' case also follows a similar pattern to the
$Z=2$ case with the lighter of $N_1^c$,
$N_2^c$ playing the role of the subdominant
right-handed neutrino in the $Z=2$ case.
It is straightforward to
scan over all the half-integer and integer charges which satisfy the
``diagonal dominated'' conditions and generate a list of charges
for this case, analagous to Table 1. A single example will suffice:
$l_3=-3/2, n_p=(0,1,2), \sigma =0$ satisfies 
all the ``diagonal dominated'' conditions and Eq.\ref{a} is saturated
by $N_2^c$ which plays the role of the subdominant right-handed
neutrino of the $Z=2$ case, with $N_1^c$ being both heavier and
having more suppressed Dirac couplings. 
Again the ``diagonal dominated'' case
can immediately be generalised to any number $Z$ of right-handed neutrinos
$n_p=(n_q, n_2,n_3)$, where $(n_2, n_3)$ are the $Z=2$ charges with 
$N_2^c$ playing the role of the subdominant right-handed neutrino
and $N_q^c$ giving subsubdominant contributions to the 23 block
of $m_{LL}$.

Examples of the ``off-diagonal dominated'' kind have
already been proposed in the literature, 
although they were not interpreted as being due to SRHND
\cite{Altarelli}. To show that the models in ref.\cite{Altarelli} are
examples of SRHND of the ``off-diagonal dominated'' kind
it suffices to consider a specific example:
\beq
l_i=(2,0,0),\ \ n_p=(1,-1,0), \ \ \sigma = 0
\label{Altcharges}
\eeq
It is immediately clear that the charges in Eq.\ref{Altcharges} satisfy
the conditions for SRHND in general Eq.\ref{modulus} and in particular 
the ``off-diagonal dominated'' conditions 
$|n_1+n_2+\sigma|<|2n_1+\sigma|,|2n_2+\sigma|$ and Eq.\ref{b}.
This immediately substantiates our claim that these models correspond to
SRHND of the ``off-diagonal dominated'' kind.
Note that $Y^{\nu}$ involves integer exponents.
In view of the interest in this example in the literature
we develop it in a little more detail below.

The charges in Eq.\ref{Altcharges}
lead to the following neutrino Yukawa and heavy Majorana matrices
\beq
Y^{\nu} \sim
\left( \begin{array}{lll}
\lambda^3 &  \lambda &  \lambda^2  \\
\lambda & \lambda & 1 \\
\lambda & \lambda & 1
\end{array}
\right), \ \ \ \ 
M_{RR} \sim
\left( \begin{array}{lll}
\lambda^2 &  1 &  \lambda  \\
1  & \lambda^2 & \lambda \\
\lambda & \lambda  & 1
\end{array}
\right) <\Sigma >
\eeq
Due to the assumed charges, the heavy Majorana matrix is dominated
by three equal mass terms 
$<\Sigma >N^c_1N^c_2$, $<\Sigma >N^c_2N^c_1$ and $<\Sigma >N^c_3N^c_3$,
leading to three roughly degenerate right-handed neutrinos.
However of the three right-handed neutrinos it is $N^c_3$ which
couples dominantly to the left-handed neutrinos of the second and
third family, due to the assumed choice of $X$ charges,
and hence dominates the 23 sector of $m_{LL}$.
To see this we evaluate $m_{LL}$ in the basis in which 
\beq
M_{RR} \sim
\left( \begin{array}{lll}
0 &  1 &  0 \\
1  & 0 & 0 \\
0 & 0  & 1
\end{array}
\right) <\Sigma >, \ \ \ \ 
M_{RR}^{-1} \sim
\left( \begin{array}{lll}
0 &  1 &  0 \\
1  & 0 & 0 \\
0 & 0  & 1
\end{array}
\right) <\Sigma >^{-1}
\label{basis}
\eeq
In this basis we define
${\tilde{Y}}_{\nu} \equiv Y_{\nu}\Omega_{RR}^{\dag}$
where
\beq
\Omega_{RR} \sim
\left( \begin{array}{lll}
1 &  \lambda^2 &  \lambda \\
\lambda^2  & 1 & \lambda \\
\lambda & \lambda  & 1
\end{array}
\right)
\eeq
Evaluating $m_{LL}$ in this basis we find from Eqs.\ref{mLL} and \ref{basis}
\beq
m_{LL}^{ij}=
\frac{v_2^2}{<\Sigma >}
(\tilde{Y}_{\nu}^{i1} {\tilde{Y}}_{\nu}^{j2 }
+ \tilde{Y}_{\nu}^{i2} {\tilde{Y}}_{\nu}^{j1 }
+\tilde{Y}_{\nu}^{i3} {\tilde{Y}}_{\nu}^{j3 })
\label{mLL '''}
\eeq
corresponding to the contributions from the inverse mass terms
$<\Sigma >N^c_1N^c_2$, $<\Sigma >N^c_2N^c_1$ and $<\Sigma >N^c_3N^c_3$,
respectively.
Since ${\tilde{Y}}_{\nu} \sim Y_{\nu}$ with the order one
contributions to ${\tilde{Y}}_{\nu}$ coming exclusively from $N^c_3$,
it is clear (by explicit evaluation of Eq.\ref{mLL '''})
that $N^c_3$ dominates the contributions
to the 23 block of $m_{LL}$, with corrections of order
$\lambda^2$ coming from the other contributions.
The remaining parts of $m_{LL}$ receive contributions at the same
order as the $N^c_3$ contributions coming from $N^c_1 , N^c_2$.
Thus the resulting light effective neutrino matrix is
as in Eq.\ref{mLL''}, with SRHND in the 23 sector due to
$N^c_3$ dominance with $O(\lambda^2)$
corrections from other right-handed neutrinos.

Finally we note that for $Z>3$ the above three categories
``diagonal dominated'', ``off-diagonal dominated'' and ``democratic''
may be combined in all possible ways.

\section{Conclusion}
We have suggested a natural explanation
of both neutrino mass hierarchies 
{\it and} large neutrino mixing angles,
as required by the atmospheric neutrino data, in terms of
a single right-handed neutrino giving the
dominant contribution to the 23 block of the light effective
neutrino matrix. We illustrated this mechanism
in the framework of models with a single pseudo-anomalous
$U(1)_X$ family symmetry, expanding all masses and
mixing angles in terms
of the Wolfenstein parameter $\lambda$.
Sub-dominant contributions to the 23 sector
from other right-handed neutrinos, suppressed by
a factor of $\lambda^2$, are required to
give small mass splittings appropriate to the small angle MSW
solution to the solar neutrino problem. 
We gave general conditions for achieving this in the framework
of $U(1)_X$ family symmetry models containing arbitrary numbers of
right-handed neutrinos $Z$. We classified the $Z=3$ cases into
three categories: ``diagonal dominated'', ``off-diagonal dominated''
and ``democratic'', and discussed examples of each kind.
Although the approach in \cite{Altarelli} 
is based on the formal condition that the subdeterminant
vanishes to order $\lambda^2$, we have shown that explicit examples 
of this kind of model may be classified within our framework as
SRHND of the ``off-diagonal dominated'' kind. Although we discussed
a particular family symmetry it is clear that 
the idea of SRHND is more general and
has recently been used in a model with $SU(2)$ family symmetry
\cite{Barbieri}.


\end{document}